\documentclass[9pt,twocolumn,twoside]{osajnl}
\journal{ao} % Choose journal (ao, aop, josaa, josab, ol, pr)

\setboolean{shortarticle}{false}
\title{Inline holographic microscopy through fiber imaging bundles}

\author[1*]{Michael R Hughes}

\affil[1]{Applied Optics Group, School of Physical Sciences, University of Kent, Canterbury. CT2 7NH. United Kingdom}

\affil[*]{Corresponding author: m.r.hughes@kent.ac.uk}

\begin{abstract}
Fiber imaging bundles are widely used as thin, passive image conduits for miniaturised and endoscopic microscopy, particularly for confocal fluorescence imaging. Holographic microscopy through fiber bundles is more challenging; phase conjugation approaches are complex and require extensive calibration. This article describes how simple inline holographic microscopy can be performed through an imaging bundle using a partially coherent illumination source from a multimode fiber. The sample is imaged in transmission, with the intensity hologram sampled by the bundle and transmitted to a remote camera. The hologram can then be numerically refocused for volumetric imaging, achieving a resolution of approximately
6~µm over a depth range of 1~mm. The scheme does not require any complex prior calibration and hence is insensitive to bending.
 
\end{abstract}

\setboolean{displaycopyright}{true}

\begin{document}

\maketitle

\section{Introduction}
Holographic microscopy is a simple yet powerful technique in which objects distributed within a sparsely occupied 3D volume can be imaged through the acquisition of a single hologram. The hologram captures phase information, providing a contrast mechanism beyond simple absorption, as well as a means of numerical refocusing. This makes it unnecessary to choose a focal plane \textit{a priori}, instead the hologram is recorded and a microscopy image can be computationally synthesised at any required depth. For samples containing multiple objects at different depths, it is possible to produce in-focus images of each object separately. This is particularly advantageous for field applications where precise focusing may be difficult \cite{zhu2013optical}, and allows for automated imaging or imaging of objects in flow.

A digital hologram is essentially a means of encoding the phase of light from the sample in the intensity pattern on the camera.  This generally requires some form of interferometry. While there are a number of different ways of achieving holographic imaging, holographic microscopy is most often performed using one of two methods. The first, known as `off-axis holography', uses a Mach-Zender type interferometer with the sample placed in one arm and the second arm acting as the reference arm \cite{cuche2000spatial}. The beam from the reference arm is arranged to reach the camera at a small tilt with respect to the sample arm beam. This creates a carrier frequency along one spatial direction of the camera, shifting the interference signal in the spatial frequency domain away from the zero spatial frequency. Band-pass filtering around this carrier and demodulating (readily achieved in the spatial frequency domain via Fast Fourier Transforms) results in a complex signal from which the phase can be extracted directly. 

In its standard form, off-axis holography requires an optical source which is both spatially and temporally coherent (even if the arms of the interferometer are exactly length-matched the tilted reference beam limits the use of very low temporal coherence sources). However, there are a number of a variations of off-axis holography that engineer a common path approach to minimise noise due to instabilities (e.g. \cite{popescu2006diffraction}) or which allow use of incoherent light \cite{guo2017off}. Some form of a reference arm is always required, and this tends to limit  miniaturisation, although some very compact implementations have been proposed for applications such as microfluidics \cite{bianco2017endowing} and endoscopy \cite{kolenovic2003miniaturized}. All variations also suffer from the disadvantage of a drastically reduced effective number of camera pixels since it is now necessary to sample the carrier spatial frequency. Off-axis holography is therefore a poor choice when camera pixels are at a premium. Non-common path system are also highly sensitive to instabilities in the optical setup or surroundings.

In the alternative `inline' approach to holography, a collimated or diverging beam is scattered by objects within the imaging volume and interferes with the unscattered portion of the beam, directly forming a hologram at the camera. Artifacts in this raw hologram can be reduced by the subtraction of a background image, acquired with no sample in the volume. The resulting `contrast hologram' can then be numerically propagated to reconstruct the intensity of objects at any or all depths within the volume. This common-path approach results in a simple and compact optical arrangement, requiring only a laser source, pinhole and camera; no other optics are required. Since there is no spatial carrier there is no loss of effective pixel count, and as inline holography is inherently common path, the setup is mechanically stable. However, quantitative recovery of phase is complicated by the twin image artifact, and requires the use of iterative algorithms with prior information such as an estimate of the spatial support of each object \cite{mudanyali2010compact} or the acquisition of multiple images \cite{greenbaum2013field}, or else more complicated optical setups such as sideband holography \cite{ramirez2015inline}. The twin image effect also introduces artefacts into the intensity image; essentially an out-of-focus copy of the image is super-imposed and, again, removal requires complicated iterative procedures.

While inline holography was traditionally performed using long coherence length lasers, a partially coherent source (such as an LED behind a pinhole) is sufficient \cite{repetto2004lensless} provided that the object-to-camera distance is kept small (typically up to a few millimeters). The small distance ensures that the scattered light hits the camera within the coherence area of the unscattered light. This simple approach has led to the development of very low cost lensless holographic microscopes, with a range of promising applications including in point-of-care diagnostics \cite{mudanyali2010compact} and water sample analysis \cite{gorocs2018deep}. Deep learning has been shown to allow quantitative recovery of phase and simulation of brightfield microscopy images \cite{rivenson2019deep}.

However, these devices all require the CCD chip to be placed very close to the sample, and so the minimum size of the microscope is governed by the requirement to include a camera and associated electronics. Despite the portability of these holographic microscopes, they tend to operate on the principle of bringing the sample to the microscope and then mounting it or flowing it through the field-of-view, much as in conventional microscopy. One can envisage an alternative use-case, in which the microscope is used as a probe and `dipped' directly into a sample, potentially allowing for on-going monitoring of the sample in its natural environment. In which case, the holographic microscopy probe would ideally be compact, passive and easy to maintain.

To this end, this article demonstrates that inline holographic microscopy can be performed using a fiber imaging bundle. Rather than capturing the inline hologram directly with a camera, the hologram is instead relayed to a remote camera. Bundles act as simple image conduits and have been widely used in endoscopic imaging applications for many years. While traditional fiberscope endoscopes have been made obsolete for most applications by the development of compact `chip-on-tip' cameras, bundles are finding new applications in endoscopic fluorescence microscopy~\cite{jabbour2012confocal}. However, a difficulty with using fiber bundles within coherent imaging systems is that even the bundles with the smallest cores (down to around 2~µm) are typically not single-mode at visible wavelengths \cite{wurster2018endoscopic}. When coherent light is employed, modal interference and coherent crosstalk between cores alters the intensity of each core, generating a speckle pattern which is highly sensitive to bending or other changes in the configuration of the fiber bundle. For example, it was noted that reflectance mode confocal microscopy through fiber bundles was inferior to spinning disk confocal using a while light source \cite{juvskattis1997real}. Optical coherence tomography through fiber bundles is also problematic, both due to modal dispersion (although some fiber bundles are single-mode in the near infra-red) and cross-talk between cores which tends to degrade interference patterns~\cite{wurster2018endoscopic}.

More significantly, the phase relationship across a field of light is not maintained by a bundle due to variations in optical path length between each of the fibers, resulting both from varying propagation constants and varying physical lengths. The optical path lengths also change as the fiber bundle is bent or otherwise disturbed. For a given configuration it is possible to measure the relative phase shift between each of the cores and correct for this using a spatial light modulator, allowing remote focusing of a spot at the distal end of the fiber, or recovery of phase at the proximal end \cite{andresen2016ultrathin}. However, this is a complex procedure, requiring recalibration for any disturbance of the fiber bundle, and has yet to find practical applications. A simple form of light field imaging can be implemented by analyzing the azimuthal dependency of power within each core \cite{orth2019optical}, but this does not allow full recovery of phase. Non-quantitative phase contrast imaging for thick tissue using oblique back illumination has also been demonstrated, but does not permit numerical refocusing \cite{ford2012phase}.

Some early attempts to demonstrate conventional holography through fiber bundles were performed using a coherent source, identifying difficulties due to the multimodal behaviour of fiber cores in the process. For example, Coquoz et al. managed to obtain reflection holographic images from Group 6 of a United States Air Force (USAF) resolution target using HeNe laser illumination delivered through a single mode fiber (SMF) \cite{coquoz1995performances}. The SMF ran parallel to the bundle and a partial reflector was placed in front of the assembly to reflect a reference beam onto the face of the bundle. To overcome multimode speckle noise within the cores, speckle was averaged using a speaker to vibrate the fiber bundle at high speed. However, no imaging from samples more realistic to potential applications was demonstrated. More recently, a conference paper from Wurster et al. demonstrated off-axis holography and demonstrated numerical refocusing through a rigid fiber bundle (a Schott image conduit) using multiple acquisitions from a wavelength swept laser both to overcome the pixel limitation and to average out speckle \cite{wurster2016lensless}. The results included a reflection image from a human finger. However, each image required 5 s to acquire which would be difficult to use with a moving bundle or sample. Furthermore, as the image conduit sat in one arm of a Mach-Zender interferometer, it is unclear if the approach would work with a flexible fiber bundle, where the phase relationship between light from each core would constantly change with bending. 

This article demonstrates transmission-mode inline holographic microscopy imaging through a fiber imaging bundle for the first time. Unlike phase conjugation approaches it does not require complex calibration and is insensitive to bending. The intensity hologram is formed at the distal end of the fiber, requiring only the intensity of the hologram to be transmitted by the bundle. Variations in optical path length between cores, therefore, do not affect the hologram or subsequent image recovery. Further, by use of a source with a short coherence length (an LED) variations in intensity due to the multimode behavior of the fiber cores are minimized; there is no need for averaging over wavelengths or vibrating the fiber. While it might be expected that cross-talk, non-regular pixelation and other degrading effects of the bundle would make holography through bundles noisy, the results demonstrate that it is in fact possible to obtain good quality inline holograms and perform numerical refocusing.

\section{Methods}

\begin{figure}[htbp]
	\centering
	\includegraphics[width=\linewidth]{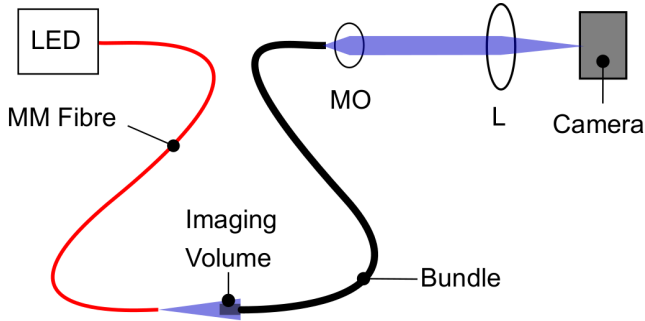}
	\caption{Setup used to assess feasibility of inline holography via fiber bundle. MO: Microscope objective lens (20X); L: tube lens, achromatic doublet (f = 100 mm); Camera: CMOS camera; MM fiber: multimode fiber for illumination; LED: 450 nm fiber-coupled LED.}
	\label{fig:schematic}
\end{figure}

A schematic of the optical set-up is shown in Fig. \ref{fig:schematic}. A 450~nm central wavelength, 15~nm bandwidth LED (Thorlabs M450LP1) was coupled into a 50~µm core, 0.22~NA multimode fiber which delivered light to the sample. The tip of the illumination fiber was approximately 15 mm from the tip of the fiber bundle (Fujikura FIGH-30-650S), and the sample was usually placed within 2~mm of the tip of the bundle. The bundle had an active imaging diameter of 600~µm, and contained approximately 30,000 cores arranged in a quasi-hexagonal pattern. The intensity of the inline hologram formed on the fiber bundle was transmitted in pixelated form and imaged onto a monochrome CMOS camera (Thorlabs DCC1545M) via a 20X objective and a 100~mm focal length tube lens. The magnification between the bundle and the camera was 11.8. With a camera pixel size of 5.2~µm, each pixel was projected to a size of approximately 0.44~µm at the bundle, and so the typical 3~µm inter-core spacing of the bundle was sampled by approximately 6 camera pixels. The image of the active area of the bundle was approximately 7.1~mm in diameter at the camera sensor plane. As the camera sensor was 6.66x5.32~mm this meant that the image of the circular bundle was slightly cropped.

\subsection{Fibre core pattern removal}

Raw images acquired through fiber bundles are corrupted by the hexagonal pattern of the fiber cores. When incoherent illumination is used with small core diameter bundles, the bundle face appears as a quasi-hexagional array of spots. The spacing between the cores - the shared cladding - appears dark. The exact size and shape of the spot from each core depends on the illumination light. Collimated light primarily couples into the fundamental mode of each core, producing smaller, Gaussian-shaped spots, while light with a random wavefront couples also into the higher order modes, resulting in a larger and potentially less regular spot. 

Two approaches for removal of the fiber core pattern were investigated. Firstly, a simple Gaussian spatial filter was applied, with a standard deviation chosen to be the minimum at which individual cores can no longer be resolved by eye in the hologram. The second approach, described in detail previously \cite{hughes2014color}, was to interpolate pixel values between the cores. This procedure requires an initial calibration using a background image. A Hough transform is used to identify the center of each of the fiber cores in the calibration image. In practice it was found necessary to first up-sample the image by a factor of 3, giving approximately 12~pixels per core diameter and 18~pixels per core spacing. A Delaunay triangulation is then formed over the core locations. A reconstruction grid is chosen (in this case corresponding to the pixels in the raw image), the enclosing triangle for each pixel in the reconstruction grid is identified, and the location of the pixel is recorded in barycentric co-ordinates. This concludes the calibration stage.

To process all subsequent holograms, the average intensity is extracted from each core in the image using the pre-calculated core position. This average is taken over the image pixels which lie inside the radius of each core, as determine by the Hough transform. This is then normalised with respect to the intensity value for this core in the calibration image. This step is designed to remove variations in core transmission as well as effects due to any small errors in locating the center of each core. The value of each pixel in the reconstruction grid, $I_p$, is then obtained by triangular linear interpolation using
\begin{equation}
I_p = I_1b_1 + I_2b_2 + I_3b_3
\label{eq:pixelRecon}
\end{equation}

where $I_n$ is the average intensity from the core lying at the $n$th vertex of the enclosing triangle, and $b_n$ is the $n$th pre-computed barycentric co-ordinate of that pixel in relation to the enclosing triangle.

\subsection{Numerical refocusing}

A contrast hologram is obtained by subtracting a background image from the raw hologram, taken with no sample in the field-of-view. This reduces spurious artifacts due to edges and variations in the intensity of the illumination source across the field-of-view. Numerical refocusing to a specific depth plane is then performed via the angular spectrum method \cite{latychevskaia2015practical}. Operationally, this is accomplished by taking a Fourier transform of the contrast hologram. This is then multiplied by the complex propagator, and the result is inverse Fourier transformed to obtain the refocused complex field. The absolute value is then taken to obtain the intensity image. Phase images can also be obtained in this way, although without further processing this is not quantitatively correct due to the presence of the twin image.

The propagator in the spatial frequency domain, $P(u,v)$ is defined as \cite{latychevskaia2015practical}
\begin{equation} 
P(u,v) = \exp \bigg [\frac{2 \pi i z}{\lambda} \sqrt{1-(\lambda u)^2 - (\lambda v)^2} \bigg]
\end{equation}

where $\lambda$ is the central wavelength, $u,v$ are spatial frequency coordinates, and $z$ is the refocus distance. Numerical refocusing of a hologram $H(x,y)$ to obtain an image $I(x,y)$ at distance $z$ is then achieved via
\begin{equation}
	I(x,y) = \mathcal{F}^{-1}[\mathcal{F}\{H(x,y)\}P(u,v)]
\end{equation}

The required refocusing depth for each image was determined automatically using a Brenner gradient-based edge detection metric to identify the best focus. This metric was found to be convex over a good range of refocus depths, allowing the use of a fast bounded search (between 100~µm and 2000~µm) using the golden section method with parabolic interpolation.

\section{Results}

\begin{figure}[htbp]
	\centering
	\includegraphics[width=\linewidth]{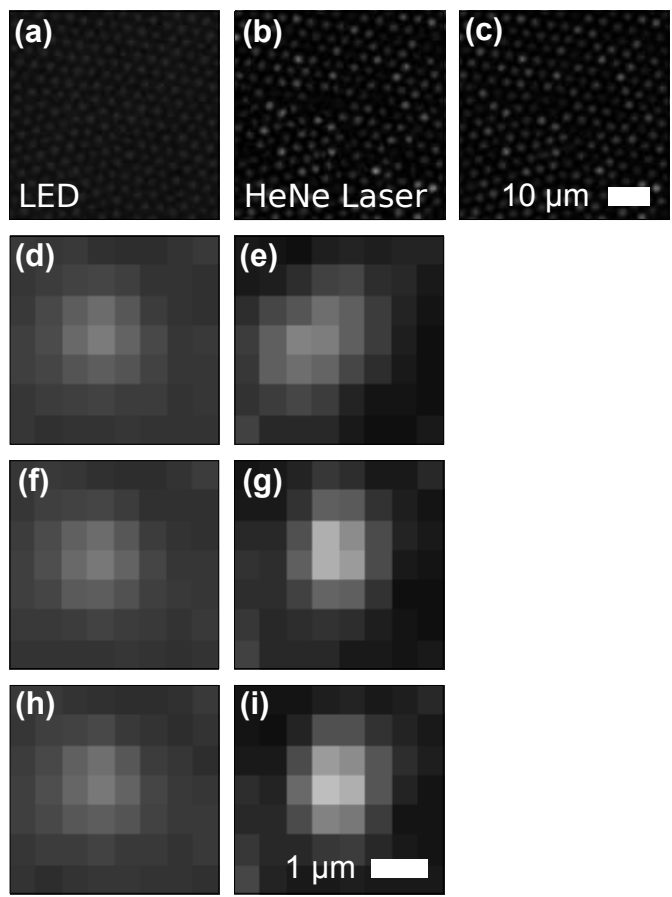}
	\caption{Demonstration of variability of core power and apparent shape depending on illumination type. (a) 50~µm x 50~µm region of fiber bundle when trans-illuminated by royal blue LED and (b) the same fiber trans-illuminated by HeNe laser. (c) is an average of 300 frames as the fiber vibrates using the HeNe laser illumination. (d,f,h) shows a single core at three timepoints for LED illumination and (e,g,i) show the same core at three timepoints for HeNe illumination.}	
	\label{fig:compare-coherence}

\end{figure}

To test the hypothesis that use of partially coherent illumination would remove random variations due to multimodal interference, the fiber bundle was illuminated by LED light via the multimode fiber with no sample present. A video of 300 frames at 30 fps was then recorded while the fiber bundle was randomly vibrated. For comparison, the experiment was repeated using illumination from a highly coherent single-mode fiber-coupled Helium Neon laser (633~nm wavelength), with the single-mode fiber directly replacing the multimode fiber used to deliver the LED illumination. Fig. \ref{fig:compare-coherence}(a) shows a zoom on a 50 x 50 µm region of an example frame from the LED illumination and (b) shows the same area of a frame from the HeNe illumination. The improved regularity of the cores and more uniform power in each core can clearly be seen in (a). Pane (c) is a mean of the 300 frames for the HeNe laser, showing that variations due to vibrations will average to produce more regular-shaped cores.

Images of a single core at three randomly chosen time points (separated by 1~s) are shown in fig. \ref{fig:compare-coherence}(d,f,h) for the LED and (e,g,i) for the HeNe laser. (The same core is shown for both illumination types). The time varying behaviour under coherent illumination can clearly be seen, whereas the shape and intensity of the core remains almost constant for the partially coherent LED illumination. It is important to note that the intensity averaged over the whole core also varies in time for the HeNe illumination (i.e. it is not possible to somehow obtain a constant spatially integrated power). To demonstrate this, the mean intensity over the core was calculated for each frame of both videos. The standard deviation divided by the mean for the LED was 0.014 while for the HeNe laser this was almost an order of magnitude greater at 0.12. This explains why it was necessary in prior work using coherent imaging through bundles (e.g. \cite{wurster2016lensless}) to average over multiple realisations of the core intensity patterns. However, as shown below, this is not necessary when using LED illumination.

\begin{figure}[htbp]
	\centering
	\includegraphics[width=\linewidth]{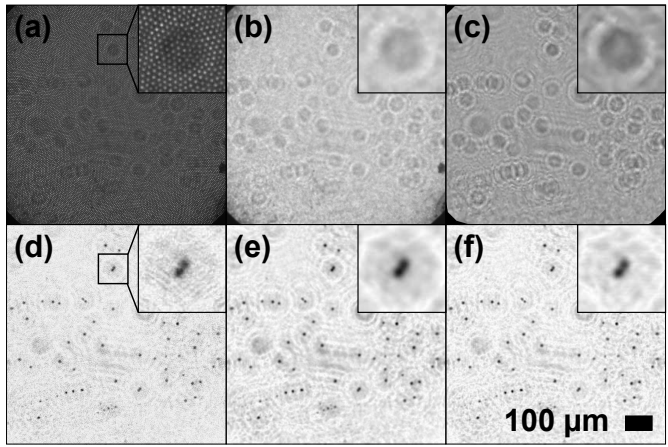}
	\caption{Inline holograms and numerically refocused reconstructions of 5~µm microspheres on glass slide acquired via fiber bundle. (a), (b) and (c) are holograms; (a) has no pre-processing, (b) has a Gaussian filter of 1.32~µm applied, (c) was linearly interpolated. (d), (e) and (f) are the corresponding numerically refocused images. Insets show a zoom on a 50x50µm area containing two closely spaced microspheres. Dataset and code available for download in Data 1 (Ref. \cite{datacode1}) }	
	\label{fig:compare-recon}

\end{figure}

Fig. \ref{fig:compare-recon} shows examples of holograms collected through the bundle using LED illumination as well as numerical refocusing of those holograms to a plane of interest. For panels (a,d) no processing was applied, for (b,e) a Gaussian filter of 3~pixels (1.32~µm) standard deviation was applied, and (c,f) used the procedure for linear interpolation between cores described above. The sample was polystyrene microspheres, with a nominal diameter of 5~µm, evaporated onto a 1~mm thick glass slide. Illumination was through the slide (i.e. there was no glass between the microspheres and the fiber bundle.) and the bundle was approximately 0.5 mm from the layer of microspheres. 

The differences between the reconstructions using the three methods are small, and the core pattern is not visible in the refocused images even when no processing is performed, although there is some high frequency noise visible in the zoomed inset. This is in contrast to contact-based imaging through fiber bundles, where the core pattern is prominent unless removed. The two microspheres in the inset are slightly better resolved for the interpolation method over the filtering, although there is scope for further optimisation of the filter to improve this. All images show examples of artifacts common in inline holography; these are discussed further in Section 4. In what follows the interpolation method is used throughout, but it should be noted that broadly similar results can be obtained  either without any processing or with simple spatial filtering.

A drawback of using fiber bundles for any kind of endoscopic imaging is the finite core or pixel count in the resulting images. The largest flexible bundles typically have around 30,000 cores, resulting in a circular image with a diameter of only around 200~pixels. In inline holography the resolution is usually limited by camera pixel size and the magnification of the interference fringe pattern onto the camera. For partially coherent sources the requirement to place the camera close to the sample means that, in practice, little better than unit magnification can be achieved. An analysis then determines the resolution to be similar to, or slightly better than, the pixel pitch \cite{mudanyali2010compact}. In the fiber bundle holographic microscope, the fiber bundle core spacing becomes the limiting factor in resolution, since the other end of the bundle can be imaged onto a camera with arbitrary magnification. A resolution comparable to the fiber core spacing is therefore expected.

\begin{figure}[phtb]
	\centering
    \includegraphics[width=\linewidth]{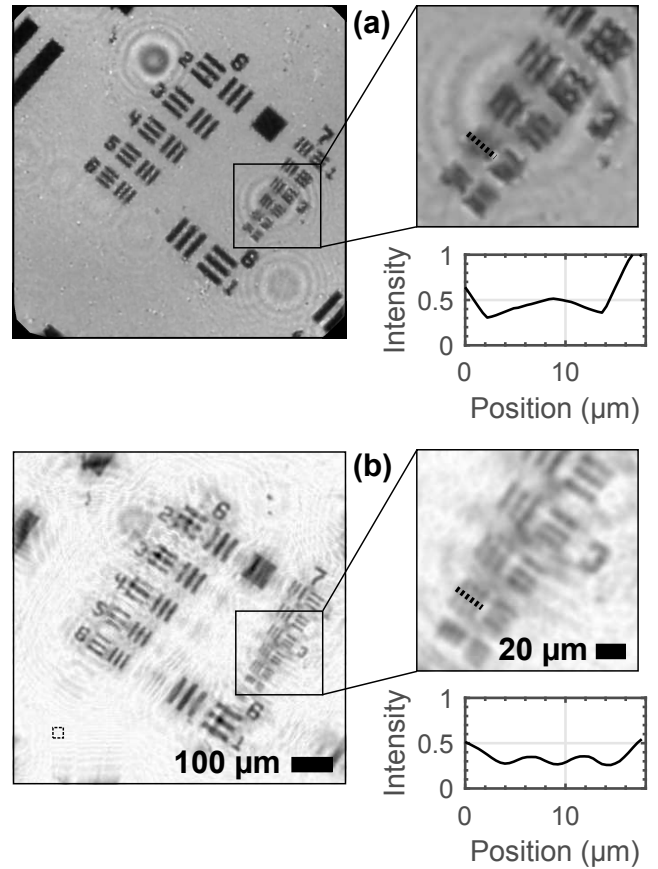}
	\caption{Images of USAF resolution target captured through fiber bundle. (a) USAF resolution target placed in direct contact with the bundle. (b) USAF resolution target at a distance of 0.75~mm from the bundle face and numerically refocused. The insets show zooms on part of Group 7. The plots show a line profile taken across Group 7 Element 5 for both (a) and (b), in the position indicated by the dashed line, with the line pattern visible in (b) but not (a). The dashed box in the lower left of the larger image in (b) shows the background area used for contrast to noise ratio calculations. Dataset and code available for download in Data 2 (Ref. \cite{datacode2})}
	\label{fig:usaf}
\end{figure}

To assess the resolution, fig. \ref{fig:usaf} shows groups 6 and 7 of a USAF resolution target. In (a) the target was placed in direct contact with the fiber bundle (i.e. no numerical refocusing was required), while in (b) it was placed at a distance of 0.75~mm from the tip of the bundle and the image was numerically refocused. Based on the typical core-core spacing of 3~µm, Nyquist sampling limits suggests a resolution of approximately 6~µm for the direct contact image, corresponding to group~7 element~3 (specified as 6.20~µm). This element is clearly resolved, while element~4 (5.52~µm) is at the borderline of visibility. For the numerically-refocused inline holography image (b), element 4 (5.52~µm) is clearly resolved and element~5 (4.92~µm) is at the limit of visibility. This can be seen most clearly in the plots showing a line profile taken through element 5. The resolution obtained through inline holography and numerical refocusing is therefore similar to, or slightly better than, conventional contact imaging.

\begin{figure}[htbp]
	\centering
	\includegraphics[width=\linewidth]{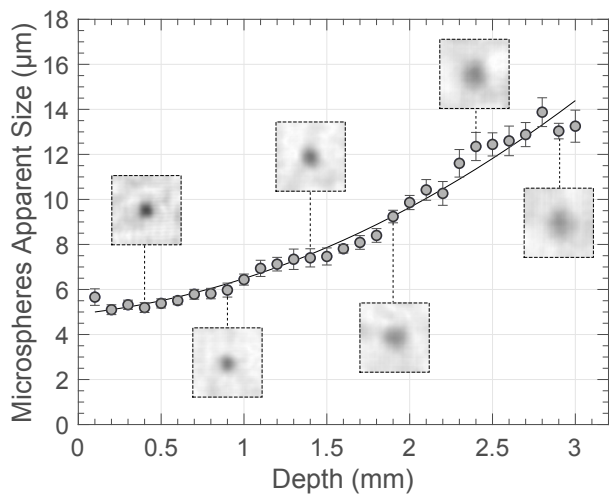}
	\caption{Effect on resolution of distance of sample from bundle face, determined by apparent size of 5~µm diameter microsphere. Values are mean full-width half-maximum across 10 spheres, error bars are standard error, trend line is 2nd order polynomial (least squares fit). Insets show example images of one of the spheres at six selected depths (27x27~µm ROI). Dataset and code available for download in Data 3 (Ref. \cite{datacode3})}
	\label{fig:resDepth}
\end{figure}

The numerically refocused image clearly contains artifacts due to the inline holography process, particularly around the larger structures. This is to be expected as the approximation of a `clean' reference field breaks down for any object of significant spatial extent. To give an indication of the severity of this, the coefficient of variation (standard deviation divided by the mean) was calculated for the pixels of a 15 x 15 µm region at the centre of the black square between the `6' and the `7'. For the contact image the coefficient of variation was 0.1 while for the refocused image it was 0.27, clearly due to the refocusing artifacts. To assess the noise away from artifacts, the contrast to noise ratio (CNR) was calculated for both images by computing the mean and standard deviation of a 15 x 15 µm background region in the bottom left of the image (away from any structures). The difference between the mean of the 15 x 15 µm region extracted from the black square and the mean of this background region was then divided by the standard deviation of the pixels within the background region, giving the CNR. The CNR was found to be 50.5 for the contact image and 34.2 for the refocused image. This suggests that the inline holography imaging leads to a small but measurable increase in noise even in the absence of obvious artifacts.

As with all inline holography systems based on partially-coherent illumination, there is a finite allowable working distance between the sample and the detector (in this case the bundle). Fig. \ref{fig:resDepth} shows the measured size of 5~µm microspheres as a function of their distance from the bundle face. Microspheres were evaporated onto a glass stage, as for fig. \ref{fig:compare-recon}, which was then moved away from the bundle using a translation stage. The best focus for each image was found using the same Brenner-based algorithm as described above, but constrained to be within 200~µm of the expected depth. To obtain the diameter of the spheres at each depth, the centre of each microsphere was taken to be the point of highest signal and the value of each pixel within a 70x70µm ROI around the centre was plotted as a function of its distance from the centre. The microsphere size was then taken to be the full-width half-maximum (FWHM) of this plot.

This measurement was repeated across 10 individual microspheres to obtain the mean FWHM values shown in the plot. The trend-line is a least-squares fit of a 2rd order polynomial ($R^2 = 0.98$). Example images of one of the microspheres at selected distances are shown for illustration. It can be seen that resolution of better than 7~µm is obtained within a distance of approximately 1~mm from the tip of the bundle, increasing to 10~µm at approximately 2~mm from the tip. This is in agreement with the result from the USAF target of a resolution of approximately 5.5~µm at a distance of 0.75~mm, shown in fig. \ref{fig:usaf}. This degradation of resolution with distance from the bundle is an expected consequence of using partially coherent illumination. The resolution also appears to degrade slightly very close to the bundle face, likely due to the impact of the fiber cores. 
  
\begin{figure}
	\centering
	\includegraphics[width=\linewidth]{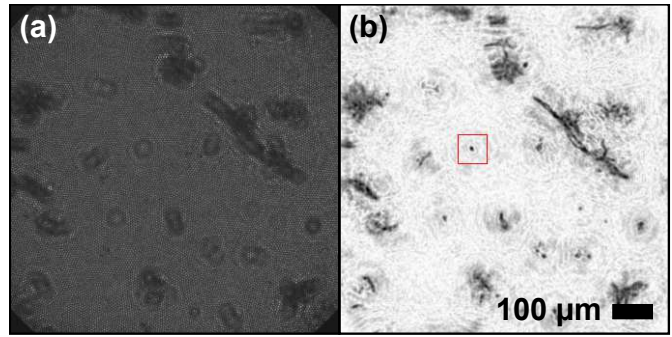}
	\caption{Numerically refocused hologram from sample of pond water. The focal position was chosen for visualisation of the object identified within the square. Video sequence showing auto-focusing on the object can be seen in Visualisation 1. Dataset and code available for download in Data 4 (Ref. \cite{datacode4}) }
	\label{fig:pond}
\end{figure}

To demonstrate the potential of the device for imaging objects distributed in 3D, the fiber was inserted into a tray of water collected from a small freshwater pond on the University of Kent campus. Fig. \ref{fig:pond} shows an example frame from Visualisation 1, a video of 130 frames acquired at 10 fps. A moving object was tracked using a simple motion identification procedure which involved subtraction of successive frames, Gaussian filtering, and identification of the resulting `peak' difference. A 55x55~µm region of interest was extracted around the moving object and the depth of the object identified using the Brenner-based algorithm described above (constrained between 0.1 and 0.5~mm). The whole frame was then numerically refocused to this depth and a 40x40~µm box drawn around the tracked object for visualization. This demonstrates the ability to individually refocus on objects of interest within the usable 3D volume of the device.

\section{Discussion and Conclusions}

The results demonstrate that is feasible to use a fiber imaging bundle to collect inline holograms and that the hologram quality is sufficient to allow microscopy intensity images to be reconstructed with resolution comparable to twice the fiber bundle core spacing.  

Since the experiments were performed using only partially-coherent light, the disadvantage of a finite depth range applies, and of course the normal limitations of the inline approach are evident. In particular, the twin artefact is clearly visible, for example as a larger circle surrounding each microsphere in fig. \ref{fig:compare-recon}. These artefacts are common to all inline holography systems and are not specific to the use of a fiber bundle. In inline holographic microscopes using fully coherent light it is possible to increase the separation between the object and the detector to a large distance; this increases the size of the twin image, spreading the power over a larger area and hence making it less visible in the intensity image. Due to the use of a partially-coherent light source here that was not possible, and so the artefact remains significant. In principle it should be possible to apply twin-artefact removal procedures applied to camera-based inline holography where real-time imaging is not required, but this remains to be investigated.

In this article only intensity image retrieval has been demonstrated. The phase information is readily available from the refocusing procedure, but this is not quantitatively correct due to the twin artefact and hence cannot reliably be used for applications such as optical thickness measurements. Further work will therefore be required to determine whether twin artifact removal for quantitative phase recovery is practical for fiber bundle holography.

A limitation of the setup used for this proof-of-concept is that the illumination fiber approaches the imaging volume from the opposite direction from the fiber bundle. The device is therefore not a 'probe' in the conventional sense as it has fibers protruding from each end, limiting the geometries into which it could be deployed. There are several ways this could be improved; for example it may be possible to route the two fibers co-axially with a reflector used to redirect the illumination light towards the fiber bundle. However, any solution must allow for the geometry required by partially-coherent illumination.

In conclusion, the results presented here demonstrate that inline holographic microscopy through fiber imaging bundles can produce high quality, high resolution numerically refocused images. This allows for holographic microscopes with an outer diameter of under 1~ mm and potentially considerably smaller, depending on the required field of view. These results may therefore open up new applications for holographic microscopy.

\vspace{10pt}
\noindent\textbf{Disclosures} 
The author declares no conflicts of interest.

% Bibliography
\bibliography{holo_endo}

% Full bibliography added automatically for Optics Letters submissions; the following line will simply be ignored if submitting to other journals.
% Note that this extra page will not count against page length
%\bibliographyfullrefs{holo_endo}

\end{document}